# Hydrodynamic interactions of low-aspect-ratio oscillating panels in a tip-to-tip formation

Yu Pan,[1,2,*] Yuanhang Zhu,[3,4] Elizabeth Westfall,[3] Daniel B. Quinn,[3] Haibo Dong,[3,*] and George V. Lauder[1,2]

[1]Department of Organismal and Evolutionary Biology, Harvard University, Cambridge, Massachusetts 02138, USA

[2]Museum of Comparative Zoology, Harvard University, Cambridge, Massachusetts, 02138, USA

[3]Department of Mechanical and Aerospace Engineering, University of Virginia, Charlottesville, Virginia 22904, USA

[4]Department of Mechanical Engineering, University of California, Riverside, California 92521, USA

*Corresponding authors: yupan@fas.havard.edu and hd6q@virginia.edu

## Abstract

The vertical, tip-to-tip arrangement of neighboring caudal fins, common in densely packed fish schools, has received much less attention than staggered or side-by-side pairings. We explore this configuration using a canonical system of two trapezoidal plates (aspect ratio $AR = 1.2$) that pitch about their leading edges while heaving harmonically at a Strouhal number $St = 0.45$ and a reduced frequency $k = 2.09$. Direct numerical simulations based on an immersed-boundary method are conducted over a Reynolds number range of $600 \leq Re \leq 1 \times 10^4$, and complementary water-channel experiments extend this range to $1 \times 10^4 \leq Re \leq 3 \times 10^4$, thereby validating the computations at higher flow speeds. Results indicate that when the plates oscillate in phase at a nondimensional vertical spacing $H/c \leq 1.0$, the cycle-averaged thrust coefficient of each plate rises by up to 14.5% relative to an isolated plate; the enhancement decreases monotonically as the spacing increases. Anti-phase motion instead lowers the time-average power coefficient by up to 6%, with only a modest thrust penalty, providing an alternative interaction regime. Flow visualization shows that in-phase kinematics accelerate the stream between the plates, intensifying the adjacent leading-edge vortices. Downstream, the initially separate vortex rings merge into a single, larger ring that is strongly compressed in the spanwise direction; this wake compression correlates with the measured thrust gain. The interaction mechanism and its quantitative benefits persist throughout the entire numerical and experimental Reynolds-number sweep, indicating weak Re-sensitivity within $600 \leq Re \leq 3 \times 10^4$. These results provide the first three-dimensional characterization of tip-to-tip flapping-plate interactions, establish scaling trends with spacing and phase, and offer a reference data set for reduced-order models of vertically stacked propulsors.

## 1. Introduction

Over their long evolutionary history, insects, birds, fish and many other organisms have evolved fast, efficient and agile locomotor abilities that are largely achieved through flapping wings or fins with diverse morphologies and oscillatory kinematics (Wu 2011). The geometry of wings and fins can be described by several metrics, among which the aspect ratio (AR) is a key parameter for characterizing their shape and determining the propulsive performance (Combes and Daniel 2001; Kruyt et al. 2015; Sambilay 1990; Usherwood and Ellington 2002; Walker and Westneat 2002). Interactions between high-AR flapping panels or foils in a tip-to-tip configuration characterized by the tips of the panels remaining closely aligned throughout the stroke cycle have been extensively studied due to interest in the hovering or forward flight dynamics of four-winged insects (Bluman and Kang 2017; Hu and Deng 2014; Liu et al. 2021; Norberg 1975; Sun and Lan 2004; Wang and Sun 2005; Wang and Russell 2007). However, flow interactions between low-AR panels in the tip-to-tip formation, such as vertically arranged fish tails shown in Fig. 1, have received limited attention.

Dense fish schools exhibit a striking range of three-dimensional sub-formations. Side-by-side, staggered, and diamond arrangements, all of which place neighboring caudal fins off the centerline, have been studied extensively for their hydrodynamic benefits. Far less attention has been paid to a conspicuous alternative: the vertical, tip-to-tip alignment of adjacent tails that emerges when fish stack above one another in crowded schools. In our recent experiments, schools of three to five giant danio (*Devario aequipinnatus*, mean length 7 cm) were filmed in a recirculating flume at 2–6 body lengths per second using synchronized high-speed cameras (Photron mini-AX50) at 250 Hz. The recordings reveal frequent episodes in which two fish occupy vertically adjacent positions, with their caudal fins aligned tip-to-tip (Figure 1 and Supplementary Movies 1–3). Despite its ubiquity in nature, the hydrodynamics of this "tail-over-tail" configuration remain largely unexplored.

The aspect ratio of propulsors exerts a first-order influence on the locomotor strategies of both flyers and swimmers, as well as on the attendant vortex dynamics. Statistical surveys place the mean AR of four-winged insects above 3.5 (Nan et al. 2018); dragonflies, for example, possess wings with AR ≈ 5 (Norberg 1975). By contrast, many fishes employ fins whose AR lies well below two (Sambilay 1990; Weihs 1989). Lake trout (*Salvelinus namaycush*) exhibit caudal fins with AR ≈ 1, and sand gobies attain values as low as 0.6 (Sambilay 1990). The AR of pectoral fins of some Labrid fish species is around 1.5 (Wainwright et al. 2002; Walker and Westneat 2002), and brook trout (*Salvelinus fontinalis*) dorsal fins have an AR of approximately 1.8 (Standen and Lauder 2007). Throughout this paper, AR is defined as $b^2/S$, where $b$ is the span and $S$ is the planform area. A high AR permits insects and birds to generate large lift with reduced lift-induced drag (Alexander 2002); a low AR, in turn, mitigates vortex-induced drag for fishes (Yeh and Alexeev 2016) and can enhance thrust production (Green and Smits 2008), acceleration (Domenici and Blake 1997; Webb 1994) and maneuverability (Flammang and Lauder 2009), while lowering bending moments that predispose fin damage (Dong et al. 2006).

Aspect ratio also determines the wake topology of three-dimensional flapping foils. At high AR, two spanwise vortices with opposite signs are shed from the trailing edge of an oscillating foil during each flapping cycle, forming a 2S wake that transitions into a reverse von Kármán vortex street when the Strouhal number is within the optimal range ($0.25 \leq St \leq 0.35$) (Triantafyllou et al. 1993). As AR decreases, streamwise vortices generated at the top and bottom panel tips are enhanced and gradually connected to the spanwise vortices (De and Sarkar 2024; Green and Smits 2008). At low AR, streamwise vortices rival the spanwise vortices in strength and dominate the evolution of the vortex wake (Buchholz and Smits 2006). Numerous experiments and simulations have shown that low-AR flapping foils generate vortex wake characterized by interconnected vortex loops, which subsequently evolve and disconnect into distinct vortex rings downstream (Buchholz and Smits 2006; 2008; Green et al. 2011). Dong et al. (2006) also observed that the vortex rings convect downstream at an angle to the streamwise direction, leading to the formation of twin oblique jets. The evolution of low-AR panel wake pattern was confirmed by digital particle image velocimetry (DPIV) measurement in live fish swimming, such as a series of linked elliptical vortex rings generated by caudal fin (AR ≈ 0.6) of chub mackerel (*Scomber japonicus*) (Nauen and Lauder 2002) and a train of separated vortex rings shed by the pectoral fin (AR ≈ 1.8) of bluegill sunfish (Drucker and Lauder 1999).

Tip-to-tip interactions between high-AR wings are well documented in four-winged insects. By adjusting the streamwise offset and phase difference between the fore- and hind-wings, these flyers manipulate wing–wake and vortex–vortex interactions to boost lift (Lehmann 2009; Xie and Huang 2015) or minimize energetic cost (Wang and Russell 2007). Equivalent studies for low-AR appendages are scarce, partly because anatomical examples are rare. Collective fish behavior, however, provides natural instances of vertically stacked, low-AR fins. A recent numerical study considered vertically aligned tuna-like bodies (Li et al. 2021), but, owing to their high-AR lunate fins (> 3), trunk–fin coupling dominated, and tail–tail coupling was weak. Species such as giant danio and trout, whose caudal fins possess AR ≤ 1, generate span-expanded vortex wakes, ideal for near-field fin–fin interaction (Menzer et al. 2025; Pan et al. 2024); yet no study has quantified the hydrodynamic consequences of aligning such tails tip-to-tip.

The present study explores hydrodynamic interactions between two identical trapezoidal panels with an aspect ratio of AR = 1.2 arranged in a tip-to-tip formation. Section 2 provides a detailed description of the panel geometry, kinematics, and the tip-to-tip configuration, together with the numerical and experimental methods. Section 3 presents immersed-boundary simulations at $Re = 600$, which sweep through vertical gap and phase offset values to map the resulting forces and efficiencies relative to an isolated panel. The accompanying flow fields are analyzed to reveal the underlying flow physics. To establish broader relevance, we extend the simulations to $Re = 2,000–10,000$ and perform water-channel experiments at to $Re = 10,000–30,000$. Collectively, these data delineate how tip-to-tip coupling reshapes the wake of low-aspect-ratio fins and how schools might exploit this configuration to modulate thrust, lift, and energetic cost.

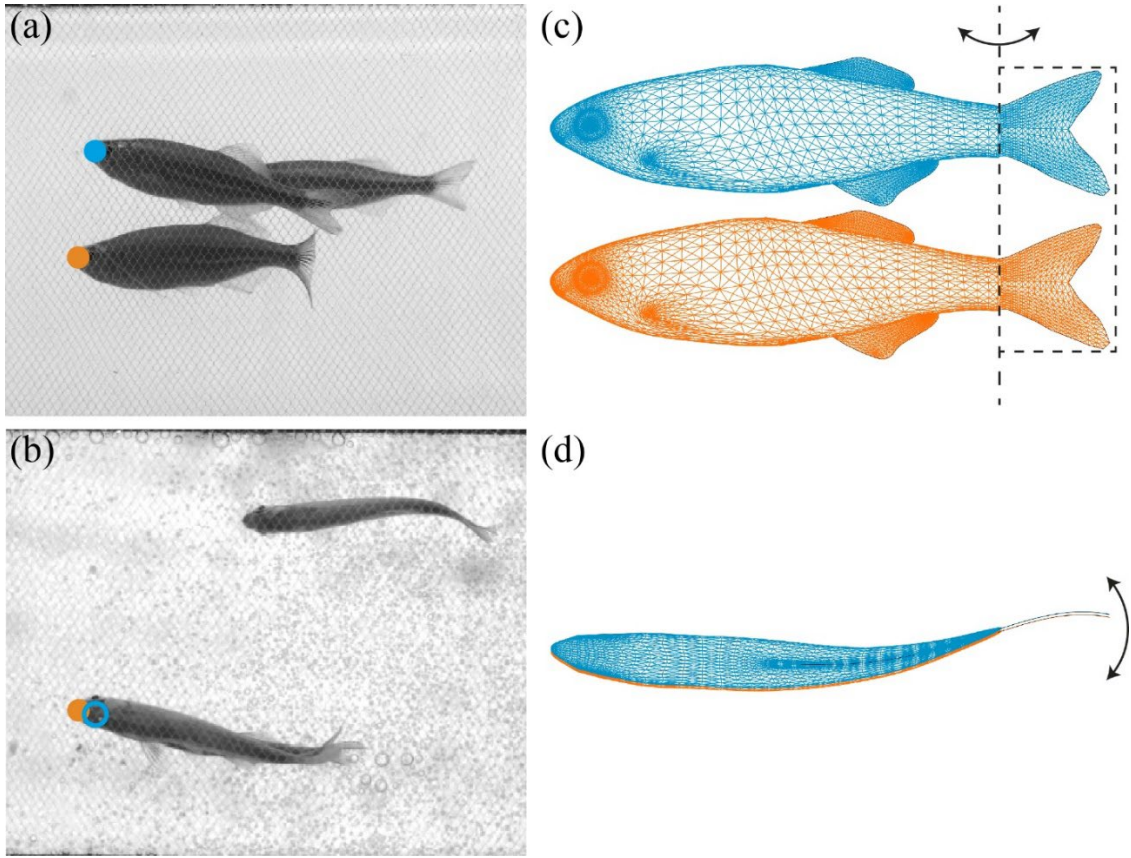

Figure 1. (a, b) Video frames from high-speed video of three giant danio swimming within a small school to show a vertical formation with the fish caudal fins in a tip-to-tip configuration at this time instant. Frames shown in (a) and (b) are from synchronized high-speed video recordings of the side and bottom views of the fish swimming against an imposed current. Vertical formations are one of several observed configurations of fish in schools (Ko et al. 2025). (c, d) Three-dimensional body models of giant danio are arranged in the same configuration. Blue and orange dots in (a) and (b) identify the two individuals swimming in a vertical tip-to-tip configuration. Also see supplementary movies 1–3.

## 2. Materials and Methods

### 2.1. Panel model, kinematics and tip-to-tip configuration

The geometry of a trapezoidal panel was modeled based on a general low-AR caudal fin shape found in various fish species where aspect ratios can approach 1.0 (Helfman et al. 1997). The relevant parameters are shown in Fig. 2(a), where $a$, $b$, $c$ represent the spans at the edges and the chord length of the panel, respectively. The AR of the trapezoidal panel is defined as the ratio of the square of the span (the longer base) to the area of the panel, i.e., $AR = b^2/S$, where $S$ is the area of the panel. In this work, the panel has a low-AR of 1.2, close to the trout caudal fin (AR = 1.3) used in a previous study (Pan et al. 2024). The panel is pitched about the leading edge and simultaneously undergoes a harmonic translational motion, i.e., a heaving motion, along the y-axis (see Fig. 2(b)). The following equations describe the motions:

$$\theta(t) = \theta_0 sin(2\pi f t + \varphi), \tag{1}$$

$$y(t) = \frac{y_0}{2} cos(2\pi f t + \varphi) \tag{2}$$

where $\theta_0 = 15°$ denotes the amplitude of the pitching motion, $y_0 = 0.4c$ is the peak-to-peak amplitude of the heaving motion, and $t$ is time. In this study, for simplicity, the pitching and heaving motions share the same frequency and phase. Thus, $f$ and $\varphi$ represent the driving frequency and phase, respectively, of the coupled motion of a panel. When two panels are vertically arranged, as shown in Fig. 2(c), $\varphi_1$ and $\varphi_2$ denote the phase of panel 1 and panel 2, respectively, and $H$ represents the tip-to-tip distance between the panels.

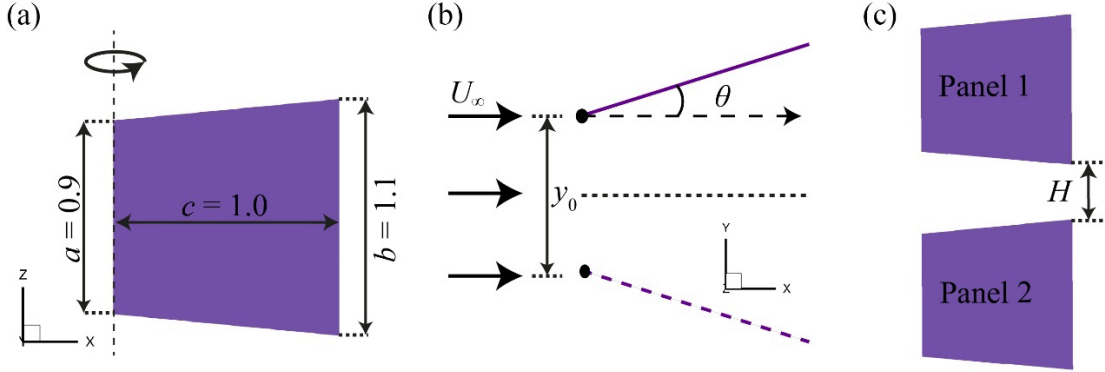

Figure 2. (a) Geometry of trapezoidal panel model, (b) top view of the pitching-heaving motion with an incoming flow, and (c) schematic of two panels in a tip-to-tip configuration separated by distance $H$.

## 2.2. Numerical methods

The governing equations of the flow problems solved are the 3D incompressible viscous Navier-Stokes equations, written in the indicial form as

$$\frac{\partial u_i}{\partial x_i} = 0; \quad \frac{\partial u_i}{\partial t} + \frac{\partial(u_i u_j)}{\partial x_j} = -\frac{1}{\rho}\frac{\partial p}{\partial x_i} + \nu \frac{\partial^2 u_i}{\partial x_j \partial x_j} \tag{3}$$

where $i, j = 1, 2$ or 3, and $u_i$ are the velocity components in the $x$-, $y$-, and $z$-directions, $p$ is the pressure, $\rho$ is the fluid density, and $\nu$ is the fluid kinematic viscosity.

Incompressible flow is computed using an immersed boundary method (IBM) based on finite-difference schemes (Mittal et al. 2008). The governing equations are spatially discretized using a collocated, cell-centered arrangement of the primitive variables, $u_i$ and $p$, and are temporally integrated via a fractional step approach. To handle the convection terms, a second-order Adams–Bashforth scheme is applied, while the viscous terms are treated implicitly using the Crank–Nicolson method to eliminate the viscous stability constraint. The simulation is constructed on non-conformal Cartesian grids, with boundary conditions accurately enforced on the immersed surfaces using a multidimensional ghost-cell strategy. We also employ a tree-topological local mesh refinement (TLMR) method on the Cartesian grids combined with parallel computing techniques to improve computational efficiency (Zhang et al. 2023) and a narrow-band level-set method for highly efficient moving boundary reconstruction (Pan et al. 2021). This numerical framework has been successfully applied to simulate oscillating propulsions (Han et al. 2022; Pan et al. 2019).

The Cartesian grid with local refinement blocks and the boundary conditions for the simulations of the pitching-heaving panels are shown in Fig. 3. The size of the computation domain is $16L \times 14L \times 14L$ with approximately 2.1 million ($145 \times 113 \times 129$) total grid points on the base layer. To accurately resolve the flow near the panels, two layers of refined meshes, represented by the red and blue blocks in Fig. 3(a), are applied, yielding the finest grid resolution of $\Delta_{min} = 0.011L$ near the panels. The zero-gradient boundary condition is applied to all lateral boundaries. The inflow boundary condition with a constant velocity is imposed at the left boundary, while the outflow boundary condition is used on the right-hand side. In addition, a grid convergence study was conducted for the pitching-heaving panel simulations. Figure 3(b) displays the time histories of the thrust coefficient of a single panel computed with four different mesh resolutions, including coarse, medium, fine, and dense meshes. The finest resolutions of these meshes are $0.035c$, $0.019c$, $0.011c$ and $0.009c$, respectively. The thrust coefficient results converge as the grid is refined. The difference in mean thrust between the fine and dense meshes is about 1.8%, indicating that the flow simulations performed on the mesh of $\Delta_{min} = 0.011L$ are grid independent.

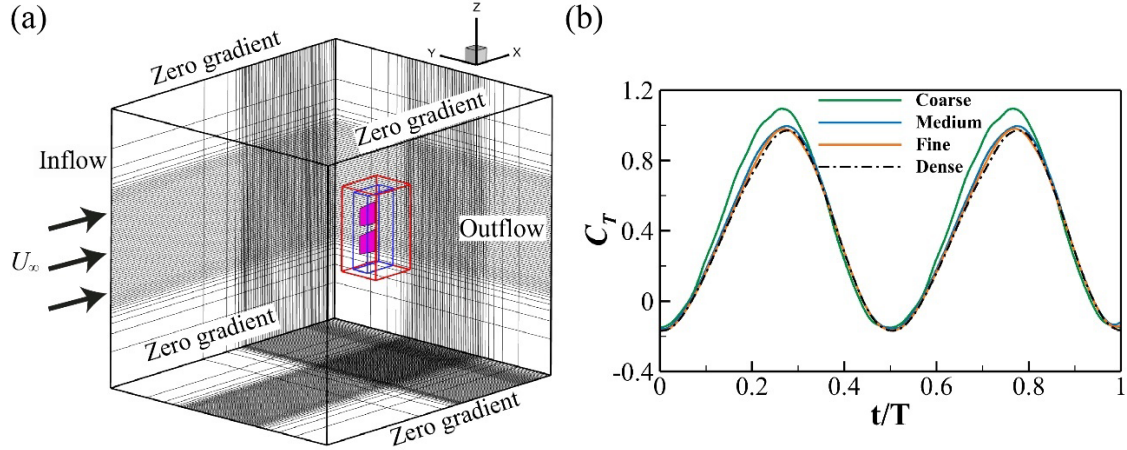

Figure 3. (a) Schematic of computational mesh and boundary conditions for the pitch-heave trapezoidal panels in a tip-to-tip configuration. (b) Comparison of the instantaneous thrust coefficients of a single pitch-heave panel computed at the coarse ($\Delta_{min} = 0.035L$), medium ($\Delta_{min} = 0.019L$), fine ($\Delta_{min} = 0.011L$) and dense ($\Delta_{min} = 0.009L$) meshes.

## 2.3. Experimental methods

We oscillated the panels in a closed-loop water channel with a test section of $0.38 \times 0.45 \times 1.52$ m ($W \times H \times L$) to cross-validate with simulation results and to extend the Reynolds number range of the present study. A schematic of the experimental setup is shown in Fig. 4. We test a single panel as the baseline case and four sets of paired panels with $H$ varied from 0 to $0.1c$, $0.25c$ and $0.5c$ to study the effect of vertical distance $H$. In the experiments, each panel has a chord length of 0.0762 m and its trapezoidal geometry matches the model used in simulations. The paired panels are connected using a carbon fiber shaft, which further attaches to a six-axis force/torque sensor (ATI Mini40 IP65) for measuring hydrodynamic forces. To reduce the influence of the carbon fiber on measurements, a teardrop cross-section is used for the panels. The pitching-heaving motion of the panels is prescribed by two servo motors (Teknic CPM-MCPV-2341S-ELN, coupled with 5:1 gearbox, SureGear PGCN23-0525), with the panel kinematics matching those used in the simulations. The motions of panels 1 and 2 are in phase, measured using a rotary encoder (US Digital EM2-1-5000-I) and a linear encoder (US Digital EM2-0-2000-I). In each experimental trial, we prescribe 50 pitch-heave cycles, with 5 additional ramp-up cycles and 5 ramp-down cycles. To increase the signal-to-noise ratio, each experimental trial is repeated 20 times for $Re = 10,000$, and 10 times each for $Re = 20,000$ and $30,000$.

We also conducted multi-layer stereoscopic PIV experiments to measure the 3D flow field (King et al. 2018; Zhu and Breuer 2023) around the paired panels. In the PIV experiments, the flow is seeded using neutrally buoyant 50 μm silver-coated hollow ceramic spheres (Potters Industries), which are illuminated by two overlapping laser sheets (L1 and L2, continuous wave, 5mm thickness). The laser sheets start from the mid-plane between the panels and are kept stationary, and we use a $z$-traverse system to raise the panels vertically in 5 mm increments, capturing the PIV layers at multiple heights along the panel. Two angled high-speed cameras (Phantom SpeedSense M341) are placed underneath the water channel to record the raw PIV images. At each layer, 750 image pairs are taken at 50 Hz. These image pairs are processed using Dantec Dynamic Studio 6.9 with an adaptive PIV algorithm (minimum interrogation window: 32 × 32 pixels; maximum: 64 × 64 pixels). Each pitch-heave cycle is divided into 30 evenly spaced bins, and the 750 two-dimensional, three-component velocity fields are phase-averaged accordingly. In total, 25 vertical layers of phase-averaged vector fields are measured. Assuming symmetry, we mirror the data about the mid-plane between the panels, forming a three-dimensional, three-component velocity field (~$4.91c \times 3.09c \times 3.15c$) that captures the entire 3D wake.

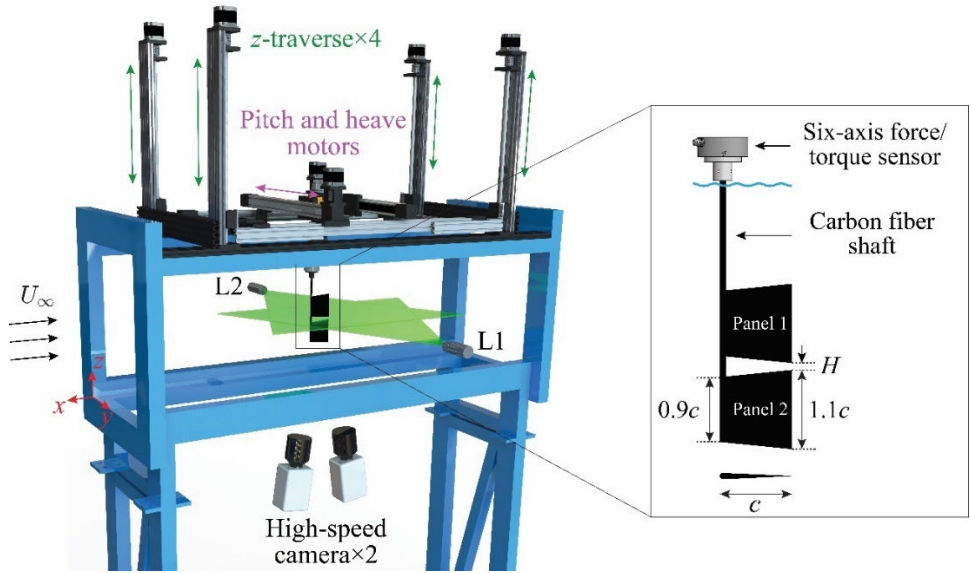

Figure 4. A schematic of the experimental setup showing a section of the recirculating flow tank with two laser light sheets (L1 and L2), two foils in a tip-to-tip configuration, and the $z$-traverse system that allows reconstruction of 3D flow fields generated by the heaving and pitching panels. The flow tank test section was $0.38 \times 0.45 \times 1.52$ m.

## 2.4 Performance measurements

The Reynolds number (Re), the Strouhal number (St) and the reduced frequency ($k$) are used to characterize the kinematics and hydrodynamics of the pitching-heaving panel propulsion. The Strouhal number is defined as the $St = fA/U_\infty$, where $A$ is the peak-to-peak trailing edge amplitude of the coupled motion, and $U_\infty$ is the incoming flow velocity, the Reynolds number is calculated by $Re = U_\infty c/\nu$, and the reduced frequency is computed as $k = \pi f c/U_\infty$. In both simulations and experiments, we utilize dynamic pressure and panel area to normalize thrust, lift and power as follows:

$$C_T = \frac{T}{0.5\rho U_\infty^2 S}, \; C_L = \frac{L}{0.5\rho U_\infty^2 S}, \; C_{PW} = \frac{P}{0.5\rho U_\infty^3 S} \tag{4}$$

where $T$ is the thrust (net force in the $x$-direction), and $L$ is the lift (net force in the $y$-direction), $P$ is the output power required to drive the panel motion and $\rho$ is fluid density. In simulations, the power $P$ is computed as $P = \oint -(\bar{\bar{\sigma}} \cdot \mathbf{n}) \cdot \mathbf{V} dS$, where $\bar{\bar{\sigma}}$ is the stress tensor, $\mathbf{V}$ is velocity, and $dS$ is the surface area element of a panel. In experiments, the power is calculated as $P = \tau \dot{\theta}$, where $\tau$ is the motor output torque measured by a transducer, and $\theta$ is the angular position of the motor measured by a rotary encoder. The measured forces and torques presented in this paper exclude the influences from the connecting carbon fiber, as well as the friction and internal stresses within the experimental setup. The propulsive efficiency of the panels is defined as $\eta = \overline{C_T}/\overline{C_{PW}}$, where overbars indicate time-averaged quantities.

## 3. Results

### 3.1. Hydrodynamic performance and wake topology of a single oscillating panel

The Strouhal number and the Reynolds number are set to 0.45 and 600, respectively, in the simulations to represent a high-thrust propulsion scenario and to replicate flow phenomena similar to those observed in fish propulsion. We maintain the Reynolds number at 600 throughout Sections 3.1–3.4 and subsequently increase it in Section 3.5 to approach conditions more closely resembling those of real fish swimming. The time histories of thrust, lift power coefficients for a single pitching-heaving panel during one cycle are displayed in Figs. 5(a) and 5(b). During one oscillation cycle, both thrust and power curves exhibit two peaks, following approximately two sinusoidal-like oscillations, while the lift curve presents a single peak with a sinusoidal-like profile. The time-averaged thrust $\overline{C_T}$ is 0.39, the mean power $\overline{C_{PW}}$ is 1.6, and the resulting propulsive efficiency $\eta$ is 0.244.

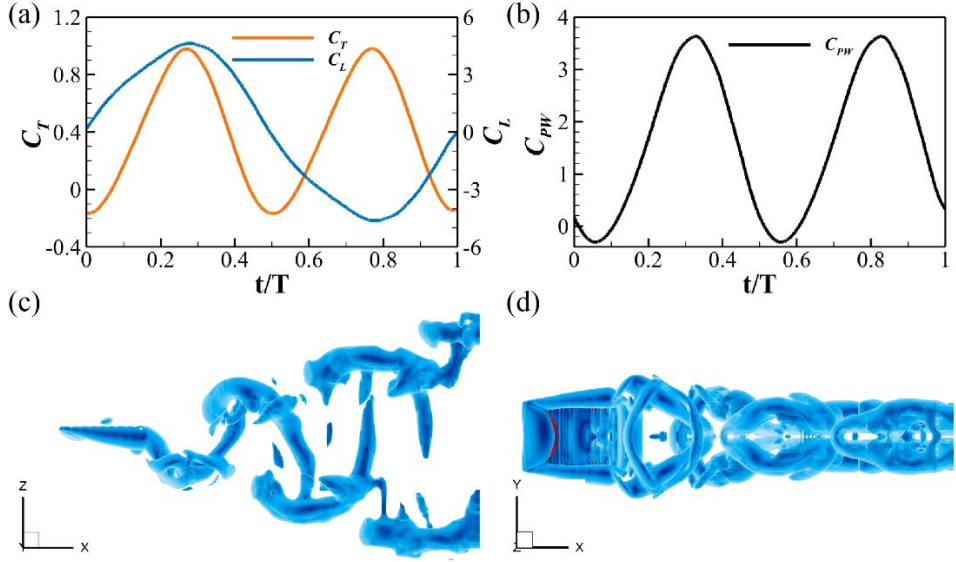

Figure 5. Computational results showing time histories of (a) thrust and lift coefficients and (b) power coefficient of the single panel during one oscillating cycle. Three-dimensional vortex structures generated by the panel at $t = 5.0T$, shown from (c) top and (d) side views. The wake structures are visualized using a dark blue iso-surface at $Q = 20$ (representing vortex cores) and a transparent blue iso-surface of $Q = 2$.

The three-dimensional vortex structures at $t = 5.0T$ are illustrated from top and side views in Figs. 5(c) and 5(d), respectively. These vortex structures are visualized by iso-surfaces of Q-criterion (Hunt et al. 1988), with a value of $Q = 20$ shown in dark blue to represent vortex cores and a values of $Q = 2$ shown in transparent blue. The Q-criterion is defined as:

$$Q = \frac{1}{2}[|\Omega|^2 - |S|^2] \tag{5}$$

where $\Omega = [\nabla u - \nabla u^T]/2$ is the vorticity tensor and $S = [\nabla u + \nabla u^T]/2$ denotes the strain rate tensor. It is observed that two rows of linked vortex rings are formed by the panel, resembling the wake structure produced by a mackerel fish (Nauen and Lauder 2002). The height of the vortex ring that is shed initially from the panel exceeds the height of its trailing edge, and the downstream vortex rings are then compressed in the spanwise direction (Fig. 5(c)). The wake dynamics of the single panel is further illustrated by the mean flow results presented in Fig. 6, which shows both 3D structures and contour plots. The 3D structures of the mean flow are visualized through the normalized velocity iso-surfaces at $\overline{U}/U_\infty = 0.94$ in blue, representing low velocity regions, and at $\overline{U}/U_\infty = 1.10$ in orange, indicating high velocity regions (Figs. 6(a) and 6(c)). Bifurcated jets form downstream of the panel (Fig. 6(a)), clearly observed in the mean flow contour plot on the horizontal slice shown in Fig. 6(b). Besides, Figs. 6(c) and 6(d) confirm that the wake compresses in the spanwise direction in the far-field region.

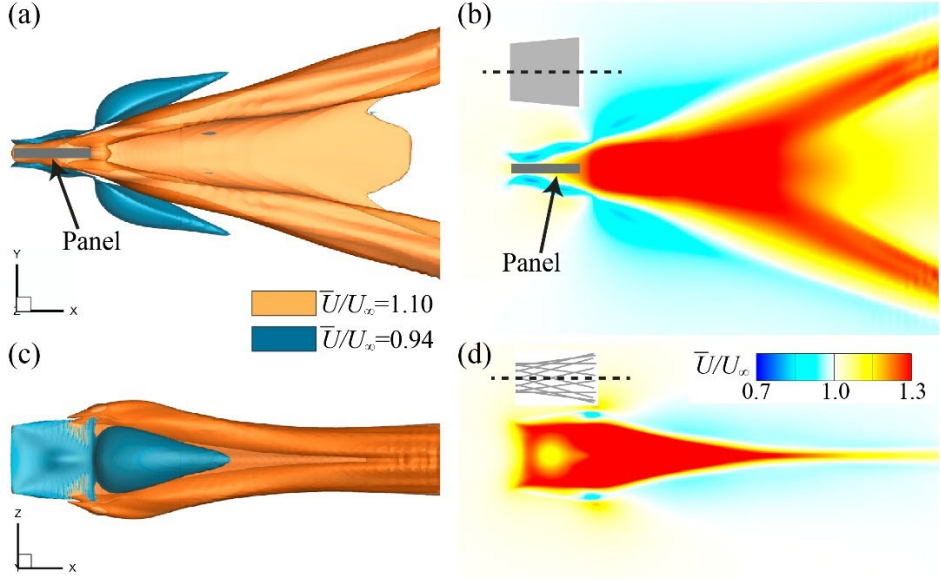

Figure 6. Computational data showing the mean flow iso-surfaces for the single panel, shown from (a) the top and (c) the side views, and contour plots of the mean flow on the (b) horizontal and (d) vertical slices.

## 3.2. Effect of vertical distance between oscillating panels in a tip-to-tip configuration

Here the effects of vertical distance $H$ on the hydrodynamic interactions between two pitching-heaving panels arranged in a tip-to-tip configuration are investigated. We vary the vertical distance $H$ from $0.1c$ to $1.0c$, while keeping panel 1 and panel 2 oscillating in-phase. Figure 7 presents the variation in hydrodynamic performance, including the thrust, power consumption and propulsive efficiency, of the panels as functions of vertical spacing, normalized by the performance of a single panel. For instance, we define $\Delta \overline{C_T}^* = (\overline{C_T} - \overline{C_{T,s}})/\overline{C_{T,s}}$, where $\overline{C_{T,s}}$ is the time-averaged thrust coefficient of the single panel. Since the two panels exhibit identical performance, only the normalized performance $\Delta \overline{C_T}^*$, $\Delta \overline{C_{PW}}^*$ and $\Delta \eta^*$ for one panel are presented. At $H = 0.1$, the thrust production for each panel in the tip-to-tip arrangement increases by 14.2% at a cost of 17.9% more power consumed compared to a single panel. This results in a slight decrease, around 3%, in propulsive efficiency. As the vertical distance $H$ increases, the interactions between panels weaken, leading to a monotonic decrease in both thrust and power consumption. Meanwhile, the normalized propulsive efficiency curve shows a V-shape trend, with $\Delta\eta^* < 0$ at all vertical distances. This indicates that in-phase tip-to-tip motion increases the thrust of the panels but reduces their propulsive efficiency compared to that of a single panel.

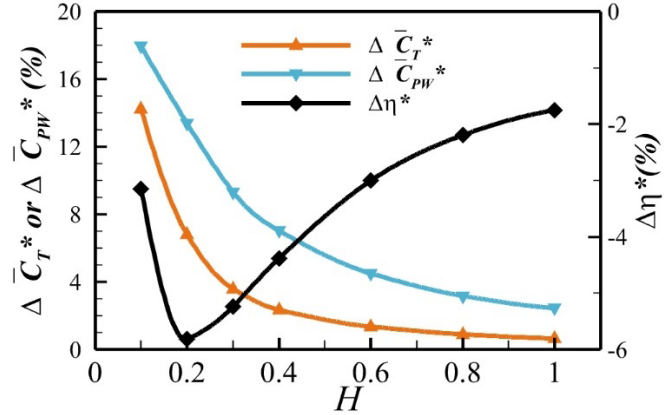

Figure 7. Computational data showing propulsive performance of panels in a tip-to-tip configuration compared to a single panel, including the normalized thrust coefficient $\Delta\overline{C_T}^*$, power coefficient $\Delta\overline{C_{PW}}^*$ and efficiency $\Delta\eta^*$.

Figure 8 shows snapshots of three-dimensional vortex structures generated by the two vertically arranged pitching-heaving panels at $H = 0.1$, captured at $t = 0.08T, 0.33T, 0.58T$ and $0.83T$, from side (a1–d1), top (a2–d2), and perspective (a3–d3) views, illustrating the evolution of the vortex wake of the panels. The vortex structures are visualized by using the same Q-criterion iso-surfaces as those in Fig. 5. Dashed lines are used to identify vortex structures, and arrows indicate the direction of flow rotation within the vortex tube. Initially, vortex rings are separately shed from each panel and subsequently merge downstream to form larger, unified rings. For example, vortex $V_{12}$, generated by panel 1, merges with the vortex $V_{22}$, shed by panel 2, forming the larger vortex $V_2$ marked by a red dashed circle in Figs. 7(c1) and 7(d1). The topology of these newly formed vortex rings evolves as they advect downstream. As depicted in Figs. 7(a1–d1), the left edge of vortex ring $V_1$ bends inward with a greater curvature than its right edges, while both the top and bottom edges bend outward, exhibiting significantly more curvature compared to that observed for a vortex ring from a single pane. Figures 7(a2–d2) illustrate the evolution of vortex rings from the top view. J-shaped individual vortex rings ($V_{12}$ and $V_{22}$) merge to form a U-shape vortex ring $V_2$ with greater curvature. Subsequently, these vortex rings compress in both spanwise and streamwise directions as they travel downstream. Compared to the single panel case, the paired panels produce larger and more strongly compressed vortex rings, which has a significant impact on the time-averaged velocity field.

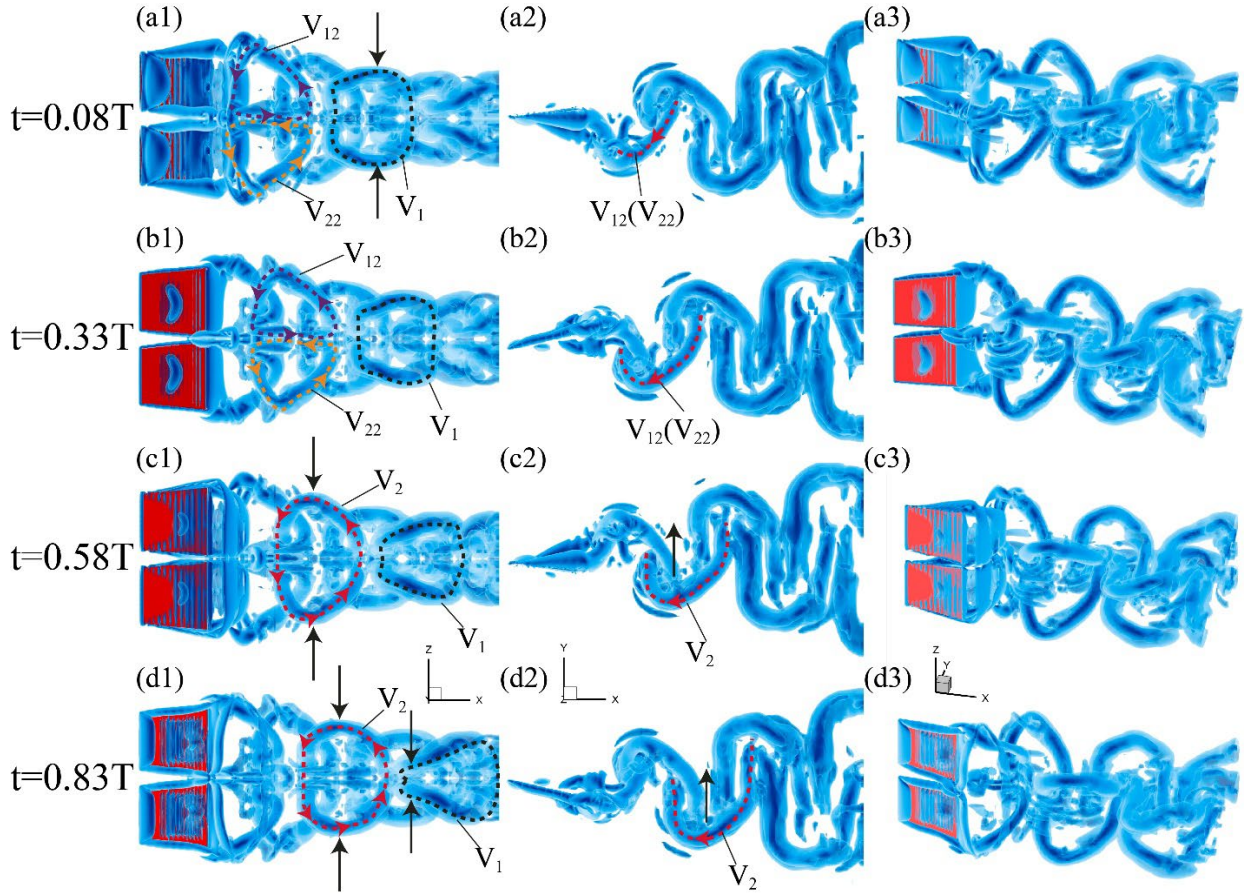

Figure 8. Computational visualization of three-dimensional wake structures generated by the panels swimming in the tip-to-tip configuration at $H = 0.1$, shown at $t = 0.08T, 0.33T, 0.58T$ and $0.83T$ from the side (a1–d1), top (a2–d2) and perspective (a3–d3) views.

Wake structures play a crucial role in determining hydrodynamic performance of swimming panels (Van Buren et al. 2017). Figure 9 shows the three-dimensional wake structures and corresponding contour plots of the time-averaged velocity field from both top and side views. Unlike the two slender jets produced by a single panel, a larger volume of high-velocity mean-flow zone is present at the center of the wake of the paired panels (Fig. 9(a)). This observation is supported by the wider high-velocity region at the center of the mean-flow contour plot on the mid-plane of the pitching-heaving motion (Fig. 9(d)). Additionally, the high-velocity zone is more compressed in the spanwise direction from the outside (Fig. 9(c)) compared to the single panel scenario. The paired panels generate a stronger and larger jet due to mutual hydrodynamic interactions, significantly increasing momentum transfer downstream and enhancing thrust production.

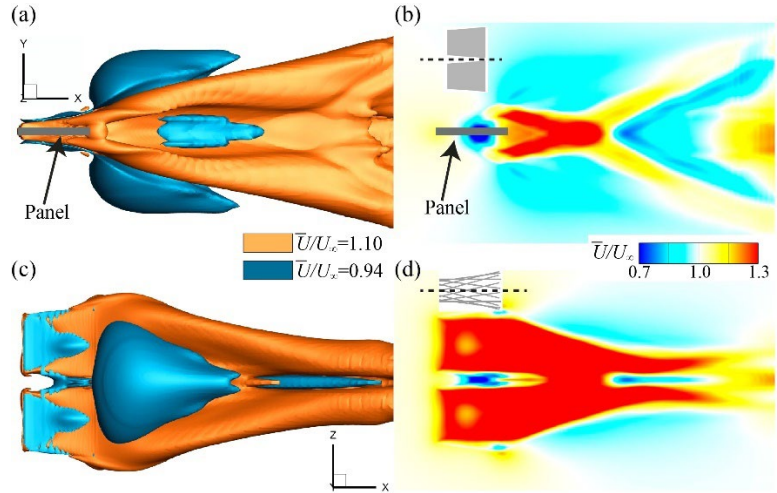

Figure 9. Computational analysis of mean flow iso-surfaces for the panels in the tip-to-tip configuration at $H = 0.1$, shown from (a) the top view and (c) the side view, and contour plots of mean streamwise velocity on (b) a slice along the horizontal mid-plane between the panels and (d) a slice through the vertical mid-plane of the pitching-heaving motion.

Figure 10 further illustrates the hydrodynamic interactions between the panels through contour plots and low-pressure iso-surfaces. The contour plots present the streamwise vorticity $\omega_x$ and normalized lateral velocity $v^* = v/U_\infty$ on a vertical slice through the center of the panels. Solid red and blue arrows indicate the panel motion directions, while dashed arrows represent flow directions. Owing to the symmetry of the pitching-heaving motion, only two representative time instances, $t = 0.08T$ and $t = 0.33T$, are depicted. In Figs. 10(a3) and 10(b3), the pressure coefficient is defined as $p^* = p/0.5\rho U_\infty^2$, where $p$ is the gauge pressure. The grey transparent outer shell presents the iso-surface at $p^* = -0.44$, while the blue inner core corresponds to $p^* = -0.89$. As the panels move, stronger leading-edge vortices (LEVs) are generated (at $t = 0.33T$) on the panels or shed ($t = 0.08T$) behind them near their mid-plane. Correspondingly, the flow velocity in the region between the panels increases, and a jet-like flow forms behind the panels, directed opposite to the panels' motion. The presence of stronger LEVs and increased flow velocity indicates enhanced thrust production of the panels. In addition, at $t = 0.08T$, low-pressure regions along the leading edges near the panels' mid-plane are connected to a large downstream low-pressure ring (Fig. 10(a3)), which is associated with the merging of separated vortex rings shed by the panels. At $t = 0.33T$, these low-pressure regions coalesce into a larger, coherent low-pressure structure shed from the trailing edges of the panels, which facilitates the transfer of flow momentum into the wake.

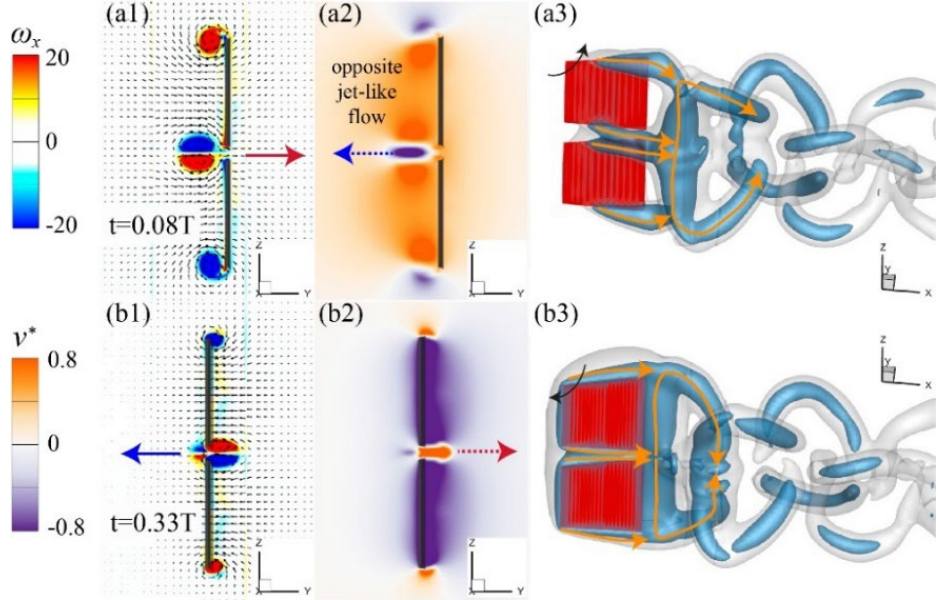

Figure 10. Computational analysis of vorticity contours $\omega_x$ (a1, b1) and normalized lateral velocity $v^*$ (a2, b2) on a vertical slice through the middle of the panels at $t = 0.08T$ and $t = 0.33T$. (a3, b3) Iso-surfaces of pressure coefficient $p^*$, defined as $p^* = p/0.5\rho U_\infty^2$, where $p$ is the gauge pressure. The transparent outer shell is visualized by $p^* = -0.44$ and the inner core by $p^* = -0.89$.

### 3.4. Effect of phase difference between oscillating panels in a tip-to-tip configuration

The panels are arranged in a tip-to-tip configuration at a vertical spacing of $H = 0.1$, and the phase of panel 1 is fixed at 0°, while the phase of panel 2 varies from 0° to 360°. The normalized propulsive performance of the panels is illustrated in Fig. 11. The trust curves in Fig. 11 show a U-shape trend for panel 1, panel 2 and their averaged value, each reaching minimum values at different phases. The thrust variation for individual panels ranges from -16% to 16%, while the average thrust of the paired panel system varies from -12% to 14%. When the phase difference is in the range $60° \leq \varphi \leq 360°$, the average thrust of the system increases compared to that of the single panel. The power consumption presents a similar trend, varying from -6% to 18% for both individual panels and the system. Notably, when the phase difference is between 90° to 270°, the power consumption of each panel is reduced relative to the single panel. Meanwhile, the propulsive efficiencies of the individual panels display an approximate sinusoidal variation with the phase difference, ranging from -13% to 4%. In contrast, the averaged efficiency presents a U-shaped curve, varying between -6% and -3%.

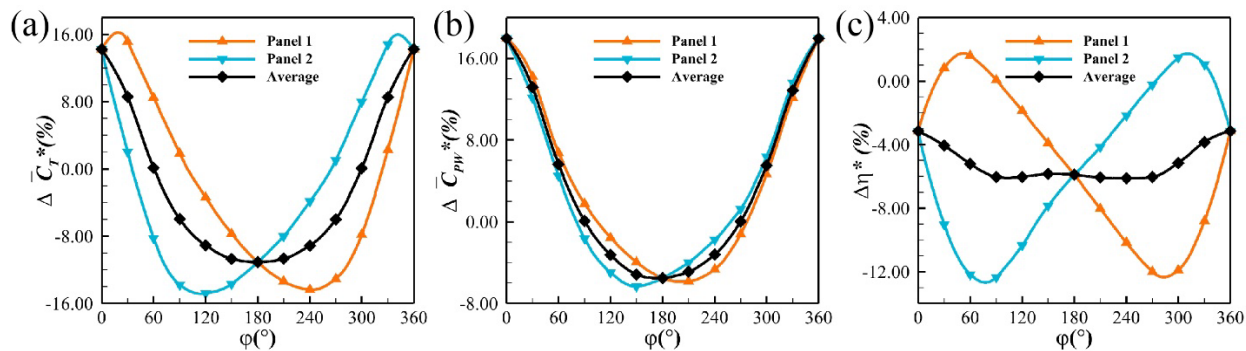

Figure 11. Computational data showing the performance variation of the panels in a tip-to-tip configuration at $H = 0.1$ compared to a single panel, including the normalized thrust coefficient $\Delta \overline{C_T}^*$, power coefficient $\Delta \overline{C_{PW}}^*$ and efficiency $\Delta \eta^*$.

At $\varphi = 180°$, the paired panel system reaches minimum values for both thrust and power consumption—thrust being 11.1% lower and power consumption, 6% lower compared to a single panel. This indicates that the anti-phase oscillation reduces the power requirement of the paired panels at the expense of reduced thrust, which is opposite to the in-phase scenario. Thus, cases with phase differences of $\varphi = 180°$ and $\varphi = 0°$ are selected for further comparison to study the effects of phase difference on hydrodynamic interactions. Figure 12 compares the three-dimensional vortex structures of the in-phase and anti-phase systems at $t = 5.0T$. The wake spreading angle $\beta$ on the horizontal plane and the narrowing angle $\alpha$ on the vertical plane for the in-phase configuration, $\beta_1 = 15°$ and $\alpha_1 = 8°$, are larger than those of the anti-phase configuration, $\beta_2 = 13$ and $\alpha_2 = 6°$. This suggests that the wake of in-phase paired panels is narrower in the vertical plane but wider horizontally. This observation aligns with findings from a single oscillating panel, where compression in the spanwise direction is coupled with spreading in the panel-normal direction (Van Buren et al. 2017).

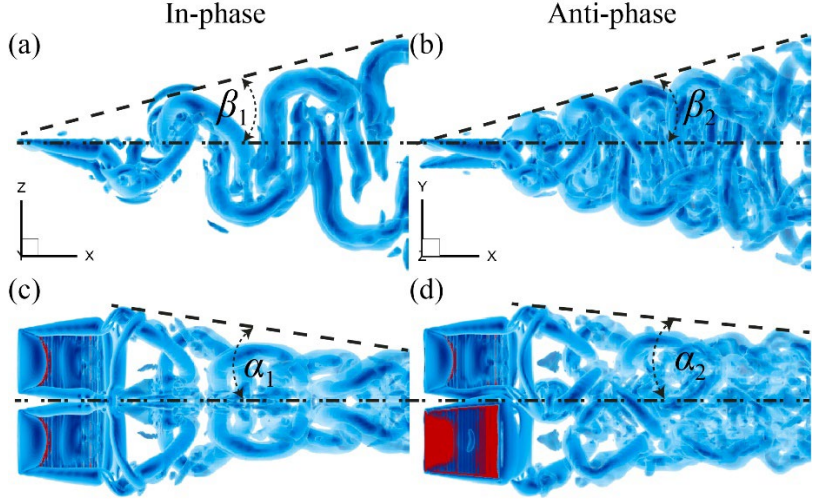

Figure 12. Computational analysis of the vortex structures resulting from in-phase ($\varphi = 0°$) (a, c) and anti-phase ($\varphi = 180°$) configurations (b, d) at $t = 5.0T$ from the top and side views.

Figure 13 compares the contours plots of streamwise vorticity $\omega_x$ and normalized lateral velocity $v^*$ on a vertical slice for the in-phase and anti-phase configurations at $t = 0.17T, 0.25T$ and $0.33T$, at three representative time instances during half an oscillation cycle. The solid arrows in Fig. 13 represent the direction of panel movement, while the dashed arrows indicate the fluid flow direction. In addition, we quantify vortex strength at these instances by calculating the circulation for a more detailed comparison. The normalized circulation $\Gamma^*$ of the LEVs is computed as:

$$\Gamma^* = \left|\frac{\Gamma}{U_\infty c}\right| = \left|\iint_{A_\Gamma} \omega_z \cdot dA_\Gamma / U_\infty c\right|, \tag{6}$$

where $A_\Gamma$ is area enclosing vorticity above or below a specific threshold value. These quantitative results, presented in Fig. 14, further substantiate the observations in Fig. 13. In Fig. 13(a1), stronger leading-edge vortices $sLEV_1$ and $sLEV_2$ have been shed by panel 1 and 2, respectively, in the in-phase configuration. When the panels move leftward, these vortices are recaptured, enhancing the flow between the panels (Fig. 13 (d1)) and strengthening the newly generated LEVs ($LEV_1$ and $LEV_2$). The enhancement continues from t=0.17T to t=0.33T, corresponding to the high-thrust generation stage of the pitching-heaving motion, significantly boosting the total thrust production. In comparison, for the anti-phase configuration, the shed vortices $sLEV_3$ and $sLEV_4$ are weaker than $sLEV_1$ and $sLEV_2$ and dissipate more rapidly, vanishing completely by t=0.33T.

Moreover, at $t = 0.33T$, the vortex $LEV_3$ induces a flow directed toward the positive y-axis along the upper leading edge of panel 2, reducing the effective velocity around the vortex $LEV_4$. A similar velocity reduction occurs near the vortex $LEV_3$ on panel 1. The weaker constructive interactions between the shed vortices and the leading edges, combined with the destructive mutual interactions between $LEV_3$ and $LEV_4$, decreases their vortex strength and consequently reduce thrust production for both panels.

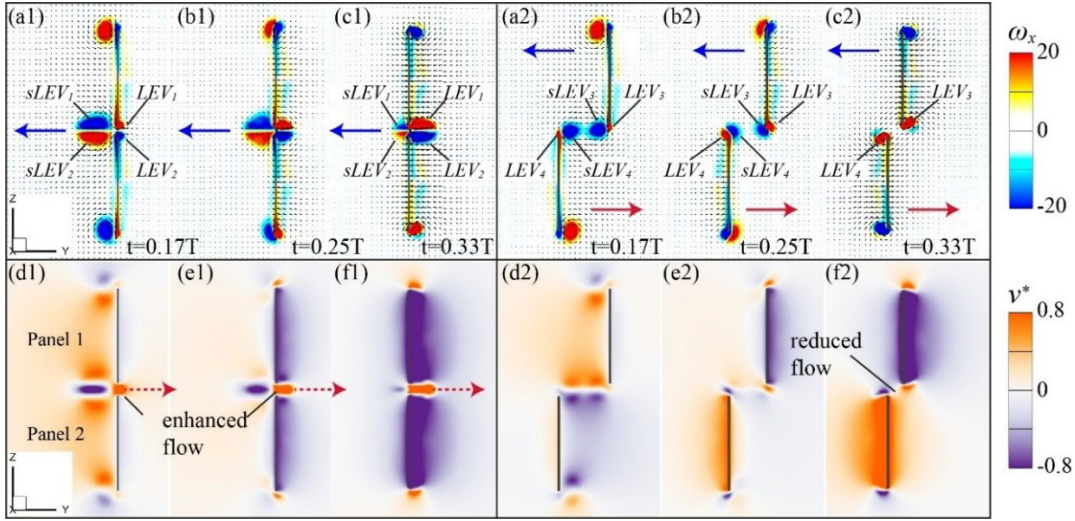

Figure 13. Computationally analysis of contour plots of vorticity $\omega_x$ (a-c) and normalized lateral velocity $v^*$ (d-f) on a vertical slice through the middle of the panels for the in-phase (a1-f1) and anti-phase (a2-f2) configurations at $t = 0.17T$ (a, d), $0.25T$ (b, e) and $0.33T$ (c, f).

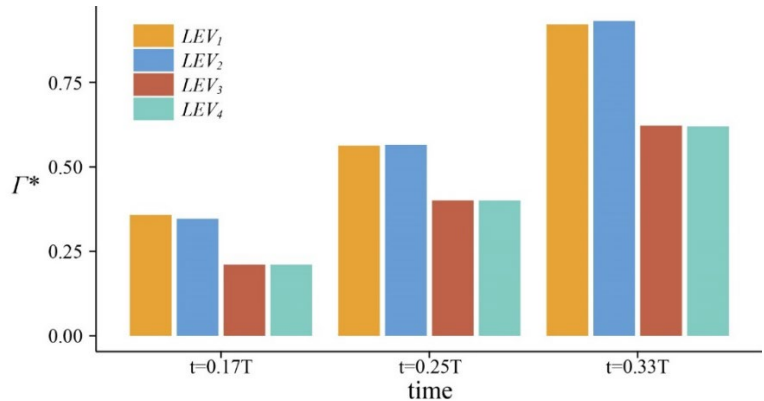

Figure 14. Normalized circulation of leading-edge vortices, including $LEV_{1,2}$ for the in-phase configuration and $LEV_{3,4}$ for the anti-phase case, at $t = 0.17T$, $0.25T$ and $0.33T$.

## 3.5. Effect of Reynolds number on thrust enhancement

In nature, fish with low-aspect-ratio fins typically swim at Reynolds numbers of $O(10^4)$ to $O(10^5)$. We therefore have expanded the Reynolds number, from 600 to $3 \times 10^5$, using both numerical simulations and water channel experiments to investigate the effects of Reynolds number on the hydrodynamic interactions between the two vertically arranged pitching-heaving panels. Computational analysis has been conducted at $Re = $ 2,000, 4,000, 6,000, and 10,000. Due to computational limitations, we performed water channel experiments to achieve higher Reynolds numbers (from $1 \times 10^4$ to $3 \times 10^4$), using the experimental setup depicted in Fig. 4.

Figure 15 compares the vortex wake structures of paired panels with a vertical distance of $H = 0.1$ at $t = 0.25T$, $0.50T$, $0.75T$ and $1.00T$ from both top and side views, obtained through experiments and CFD simulations, both at $Re = 1 \times 10^4$. Figures 15(a1-a4) and 15(c1-c4) present the wake observed from experiments, while Figs. 15(b1-b4) and 15(d1-d4) show numerically obtained vortex wakes. CFD results and experimental results are both visualized by using iso-surface of Q-criterion iso-surface at 2% of $Q_{max}$, colored by the spanwise vorticity $\omega_z$. Vortex rings shed from the panel trailing edges interact and form curved, hook-like vortex chains, as shown in the top views for both simulations (Figs. 15(a1-a4)) and experiments (Figs. 15(b1-b4)). Additionally, the side views show that the wakes are compressed in the spanwise direction in both experiments and simulations. The plot suggests that DNS captures the main unsteady flow features observed experimentally at high Reynolds numbers accurately.

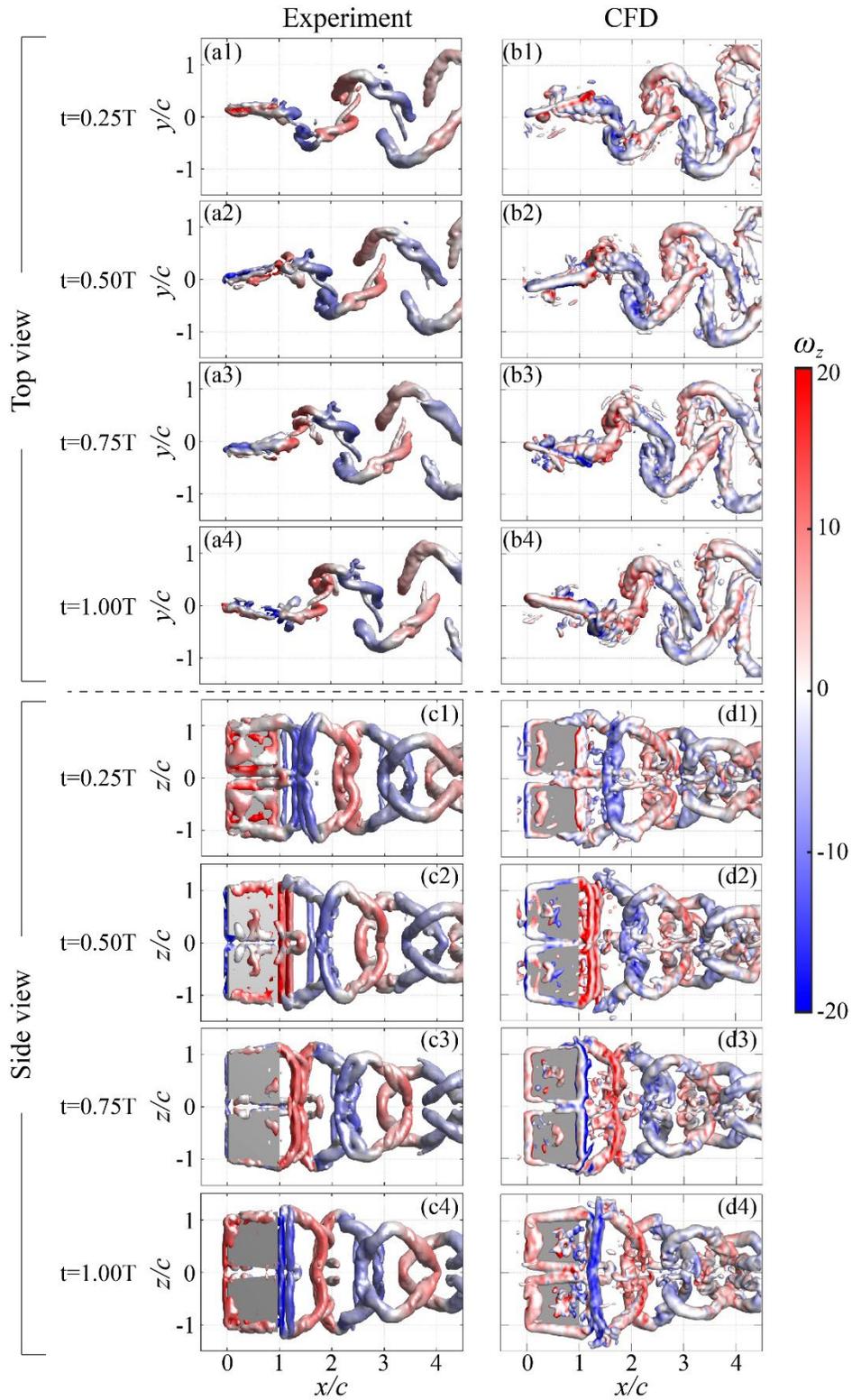

Figure 15. Comparison of vortex flow structures resulting from experiments (a, c) and CFD simulations (b, d) at $Re = 1 \times 10^4$ at $t = 0.25T$, $0.50T$, $0.75T$ and $1.00T$. Also see supplementary movies 4–5.

Figure 16 presents the time histories of the thrust coefficient ($C_T$), lift coefficient ($C_L$) and power consumption coefficient ($C_{PW}$) of the paired panels at different Reynolds numbers and their variation for both experiments and simulations. Due to symmetry and in-phase motion, the two panels exhibit identical hydrodynamic performance; therefore, results for only one panel are shown. In Fig. 16(a), the simulated $C_T$ closely matches the experimental measurement at $Re = 1 \times 10^4$. The time-averaged thrust coefficient from simulation is $\overline{C_{T,CFD}} = 0.682$, while the experimental result is $\overline{C_{T,EXP}} = 0.641$. The minor difference (~6%) confirms the high accuracy of our DNS solver. In Fig. 16(b), the peak of the simulated lift coefficient slightly lags behind the experimental peak, with a marginally larger amplitude for numerical results. Similar minor deviations between simulations and experiments appear in the $C_{PW}$ curves. Such discrepancies between simulations and experiments are minimal and acceptable, especially for high Reynolds number flows. Possible reasons include geometric differences: zero-thickness membranes were used in simulations, whereas hydrofoils with teardrop-shaped cross-section were employed in experiments. Furthermore, experimental setup, such as the rods connecting the two panels (see Fig. 16(d)) and water channel walls, might introduce additional performance variations.

Fig. 16(d) presents the variation in thrust, power consumption and propulsive efficiency relative to a single panel at $Re = 1 \times 10^4$ in the experiments with the vertical distance $H$ varying from 0 to 0.5. In Fig. 16(d), the thrust ($\Delta \overline{C_T}^*$) and power consumption ($\Delta \overline{C_{PW}}^*$) of the paired panels increases with decreasing $H$, consistent with trends observed at $Re = 600$ (see Fig. 7). Figure 16(e) displays the thrust variation of the paired panels at a fixed vertical distance ($H = 0.1$) across different Reynolds numbers, based on both experimental and simulation results. Although the thrust enhancement for the panels at high Re is lower than that at medium Re, its minimum value remains around 10% across the entire range. The trend shown by fit curve suggests that thrust enhancement rises at even higher Reynolds number. Both numerical and experimental results confirm that significant thrust enhancement occurs for vertically arranged pitching-heaving panels across a wide range of Reynolds numbers.

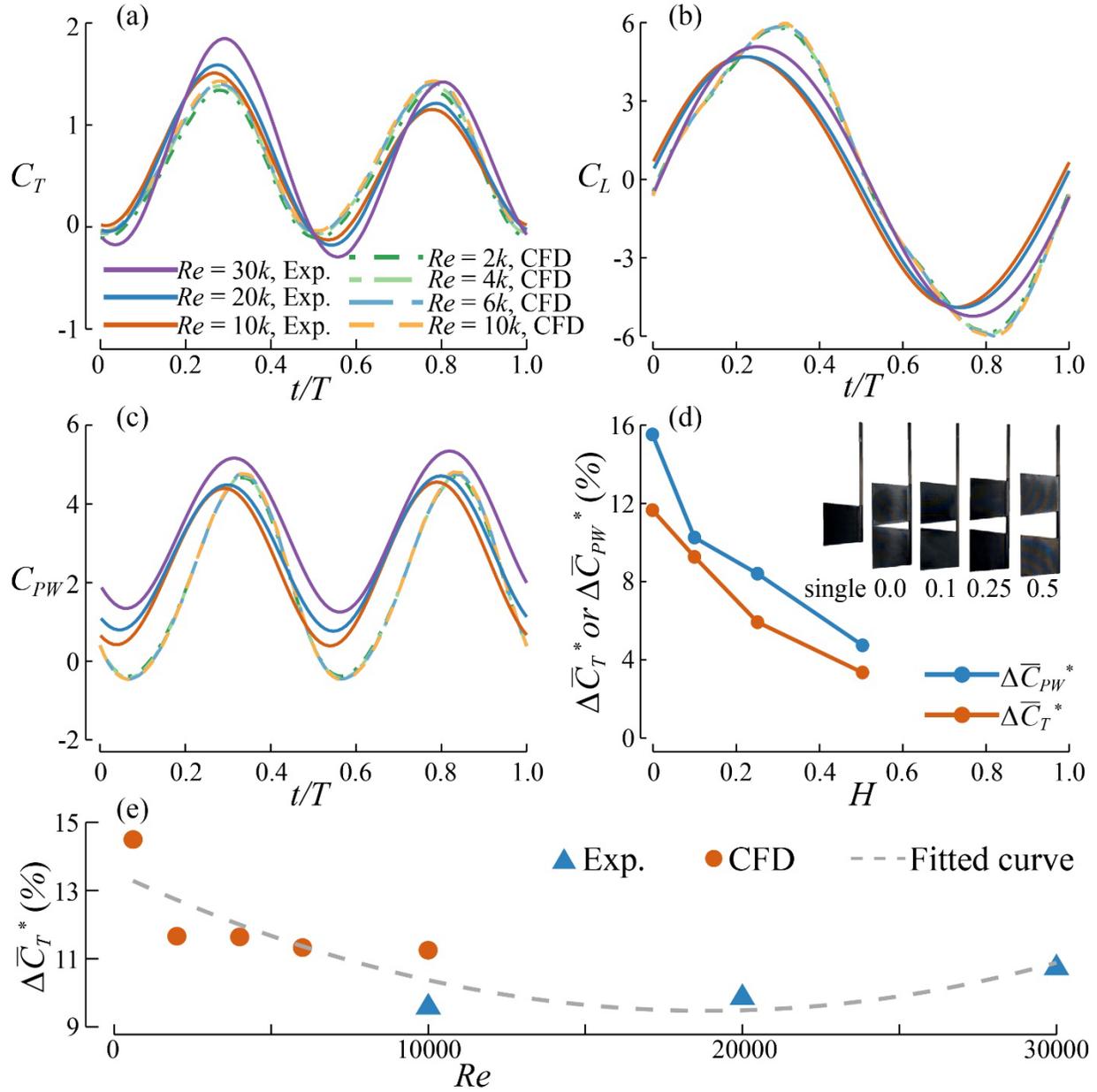

Figure 16. Time histories of (a) thrust ($C_T$), (b) lift ($C_L$), and (c) power consumption ($C_{PW}$) coefficients of in-phase paired panels oscillating at high Reynolds number from experiments ($Re = 1 \times 10^4$, $2 \times 10^4$ and $3 \times 10^4$) and simulations ($Re = 2{,}000$, $4{,}000$, $6{,}000$ and $1 \times 10^4$). (d) At $Re = 1 \times 10^4$, the normalized variation of time-averaged thrust and power consumption of paired panels, $\Delta \overline{C_T}^*$ and $\Delta \overline{C_{PW}}^*$, are shown as functions of the vertical distance $H$ in experiments. (e) Thrust variation $\Delta \overline{C_T}^*$ of the in-phase paired panels with a vertical distance of $H = 0.1$ from simulations and experiments across a broad range of Reynolds number. The grey dashed line represents the fitted curve of thrust variation versus Reynolds number.

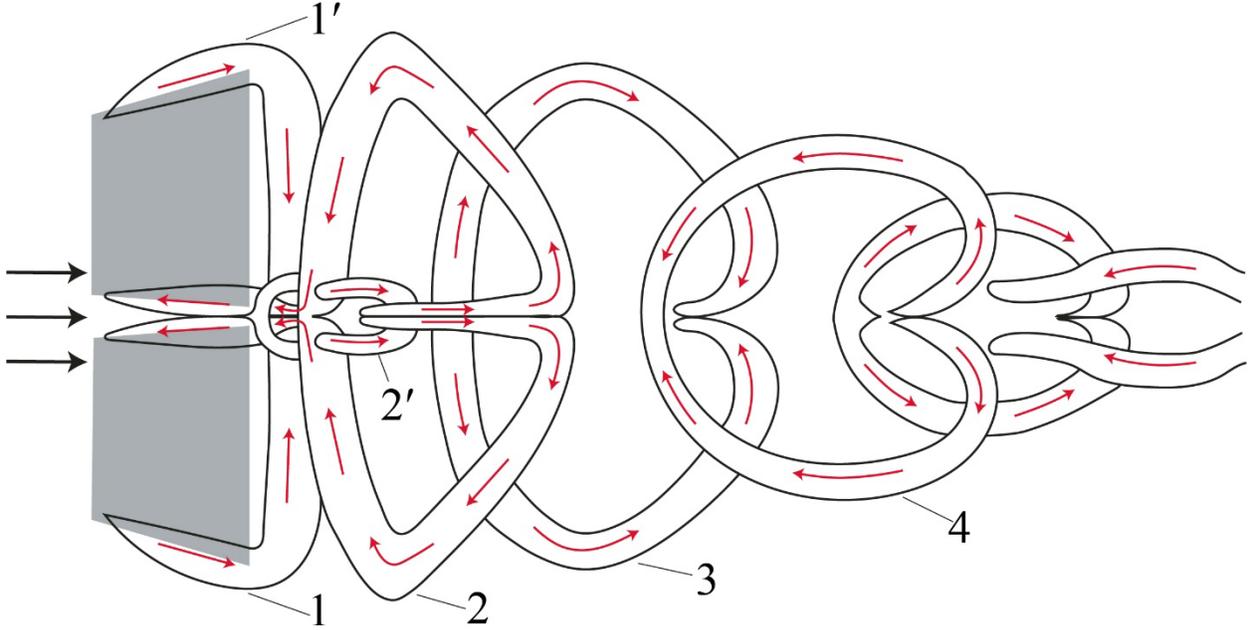

Figure 17. Schematic of the wake generated by vertically arranged pitching-heaving panels. Vortex rings 1 and 1' are initially shed separately from each panel, then merge into a larger vortex ring, labeled vortex 2, accompanied by a secondary horseshoe-shaped vortex structure, vortex 2'. As the vortex rings are transported downstream, they become compressed in the vertical direction. Red arrows indicate the rotational direction of the vortex tubes, while black arrows represent the direction of the incoming flow.

## 4. Discussion and conclusion

This study provides the first three-dimensional characterization of hydrodynamic interactions between two low-aspect-ratio propulsors arranged tip-to-tip, a vertical alignment frequently observed when fish stack above one another in dense schools. We examined a canonical pair of rigid trapezoidal plates (AR = 1.2) that pitch about their leading edges while heaving harmonically at a Strouhal number of 0.45 and a reduced frequency of 2.09. Direct numerical simulations, performed using an immersed-boundary method, covered the range $600 \leq Re \leq 1 \times 10^4$, and complementary water-channel experiments extended the range to $3 \times 10^4$, allowing for a combined numerical–experimental data set that spans over two orders of magnitude in swimming speeds.

Numerical simulations at $Re = 600$ suggest that in-phase oscillation enhances the thrust production of each panel by up to 14.5% (at $H = 0.1c$), though at the cost of a minor reduction (~3%) in propulsive efficiency. Results from both experiments and simulations indicate that the thrust enhancement remains significant at high Reynolds up to $Re = 3 \times 10^4$. Anti-phase oscillation undermines the thrust enhancement, but with the benefit of reducing power consumption by up to 6%. The hydrodynamic interactions weaken as the vertical distance increases, and when $H \geq 0.5c$, the interactions are minimal.

Considering the shape and dynamics of oscillating panels and the resulting wake patterns, hydrodynamic interactions among panels arranged in a planar configuration can significantly enhance swimming performance. Under 2D assumptions—i.e., when the aspect ratio is large—the average thrust of panels can increase by up to 35% when they oscillate in an in-line formation (Boschitsch et al. 2014) or in a side-by-side formation (Dewey et al. 2014). Staggered formation can boost the thrust enhancement of the system up to 87% (Ormonde et al. 2024). However, three-dimensional effects introduced by decreasing AR strengthen the interactions between the tip vortex and trailing-edge vortex, greatly altering the performance and vortex dynamics of panels (Buchholz and Smits 2006; 2008; Green and Smits 2008). This variation significantly reduces the performance gains of collective swimming in planar configurations. At low AR, the thrust enhancement in side-by-side and staggered formations is reduced to around 10% (Kurt and Moored 2018) and 34% (Kurt et al. 2020), respectively. In in-line formations, three-dimensional effects can even turn constructive interactions at high AR into destructive interactions at low AR (Han et al. 2022). The deterioration of hydrodynamic interactions between three-dimensional oscillating panels arranged in planar configurations raises an important question: do interactions occur between panels arranged in 3D space? Our results confirm that strong hydrodynamic interactions exist between vertically spaced panels and suggest that such configurations can significantly boost thrust production. The observed performance enhancement is comparable to that seen in other configurations and should not be overlooked. It is also important to note that, for simplicity, this study employed rigid trapezoidal panels oscillating with synchronized pitching-heaving motion and a fixed Strouhal number. Nevertheless, studies indicate that these parameters determine performance and vortex structures of the panels (30, 50, 55, 56), and they may substantially affect inter-panel interactions, potentially further enhancing performance in a tip-to-tip configuration. Additionally, the trend in the fitted curve in Fig. 16(e) suggests that thrust enhancement may continue to increase at higher Reynolds numbers.

Flow visualization reveals that thrust enhancement in the in-phase configuration is attributed to a flow increase in between the paired panels and recapture of strengthened shed vortices, leading to stronger LEVs generated on the adjacent leading-edges. These enhancements coincide with notable variation in downstream wake structures. Figure 17 provides a schematic illustrating vortex merging and wake evolution for the paired panels. Initially, in Fig. 17, the vortex rings, vortex 1 and 1', are shed separately from the trailing edges of each panel, and the adjacent vortex tubes begin merging due to their close proximity. After one oscillating cycle, the separated vortex wings coalesce into a larger vortex ring, vortex 2, accompanied by a secondary horseshoe-shaped vortex structure labeled vortex 2'. Vortex 2 initially maintains a height equivalent to the combined heights of the original separated vortex rings. However, as it moves downstream, vortex 2 undergoes significant vertical compression, which is typically an indicator of high thrust generation in pitching-heaving systems (Van Buren et al. 2017). Moreover, vortex wake measured from both simulations and experiments at a Reynolds number of $Re = 1 \times 10^4$ in Fig. 15 demonstrate that the physical mechanism responsible for thrust enhancement at intermediate Re remains consistent at high Re. As a comparison, during the anti-phase motion, mutual interactions between the LEVs decrease the effective velocity around their respective leading edges and weaken each other's vortex strength, which subsequently undermines wake recapture. Although this reduces thrust, the decreased effective velocity results in lower power consumption. Additionally, the wake produced by the anti-phase configuration is narrower in the horizontal plane but wider in the vertical plane than that of the in-phase configuration. This observation aligns with previous findings for oscillating panel systems: spanwise compression of the wake is associated with thrust enhancement, while it is also coupled with wake spreading in the panel-normal direction (Van Buren et al. 2017).

These findings show that tip-to-tip stacking can deliver thrust gains or energetic savings comparable to those reported for staggered or side-by-side arrangements, suggesting that vertically aligned caudal fins represent a viable hydrodynamic strategy for fish in dense schools. The dataset established here provides a benchmark for reduced-order models of vertically stacked propulsors and points toward design rules for multiple underwater vehicles that must balance thrust and efficiency. Future work should extend the present framework to compliant fins, asynchronous pitching–heaving phases, and varied Strouhal number regimes to determine whether even greater synergistic gains are achievable and to map the stability boundaries of stacked arrays under biologically relevant conditions.

## Acknowledgements

This work was supported by the Office of Naval Research (ONR) [grants N00014-21-1-2661, N00014-16-1-2515, N00014-15-1-2234 and 00014-22-1-2616] and the National Science Foundation (NSF) [grants CNS-1931929 and EFRI-1830881].

## Data accessibility

All data generated are available in the paper or its electronic supplementary material.

## Declaration of interests

The authors report no conflict of interest.

## Supplementary movies

Supplementary movies s will be made available upon request or through the publisher's platform upon acceptance.

## References

**Alexander DE** (2002) *Nature's flyers: birds, insects, and the biomechanics of flight*. JHU Press.
**Bluman J and Kang CK** (2017) Wing-wake interaction destabilizes hover equilibrium of a flapping insect-scale wing. *Bioinspir Biomim* **12**(4)**,** 046004. https://doi.org/10.1088/1748-3190/aa7085.
**Boschitsch BM, Dewey PA and Smits AJ** (2014) Propulsive performance of unsteady tandem hydrofoils in an in-line configuration. *Physics of Fluids* **26**(5 ).
**Buchholz JHJ and Smits AJ** (2006) On the evolution of the wake structure produced by a low-aspect-ratio pitching panel. *Journal of Fluid Mechanics* **546,** 433-443.
**Buchholz JHJ and Smits AJ** (2008) The wake structure and thrust performance of a rigid low-aspect-ratio pitching panel. *Journal of Fluid Mechanics* **603,** 331-365.
**Combes SA and Daniel TL** (2001) Shape, flapping and flexion: wing and fin design for forward flight. *The Journal of Experimental Biology* **204**(Pt 12)**,** 2073-2085. https://doi.org/10.1242/jeb.204.12.2073.
**De AK and Sarkar S** (2024) Evolution of wake structure with aspect ratio behind a thin pitching panel. *Journal of Fluids and Structures* **124,** 104025.
**Dewey PA, Quinn DB, Boschitsch BM and Smits AJ** (2014) Propulsive performance of unsteady tandem hydrofoils in a side-by-side configuration. *Physics of Fluids* **26**(4).
**Domenici P and Blake RW** (1997) The kinematics and performance of fish fast-start swimming. *Journal of Experimental Biology* **200**(8)**,** 1165-1178.
**Dong H, Mittal R and Najjar FM** (2006) Wake topology and hydrodynamic performance of low-aspect-ratio flapping foils. *Journal of Fluid Mechanics* **566,** 309-343.
**Drucker EG and Lauder GV** (1999) Locomotor forces on a swimming fish: three-dimensional vortex wake dynamics quantified using digital particle image velocimetry. *Journal of Experimental Biology* **202**(18)**,** 2393-2412.

**Flammang BE and Lauder GV** (2009) Caudal fin shape modulation and control during acceleration, braking and backing maneuvers in bluegill sunfish, Lepomis macrochirus. *Journal of Experimental Biology* **212**(2)**,** 277-286.

**Green MA, Rowley CW and Smits AJ** (2011) The unsteady three-dimensional wake produced by a trapezoidal pitching panel. *Journal of Fluid Mechanics* **685,** 117-145.

**Green MA and Smits AJ** (2008) Effects of three-dimensionality on thrust production by a pitching panel. *Journal of Fluid Mechanics* **615,** 211-220.

**Han P, Pan Y, Liu G and Dong H** (2022) Propulsive performance and vortex wakes of multiple tandem foils pitching in-line. *Journal of Fluids and Structures* **108,** 103422.

**Helfman GS, Collette BB and Facey DE** (1997) The Diversity of Fishes Malden, Massachusetts. In.: Blackwell Science.

**Hu Z and Deng X-Y** (2014) Aerodynamic interaction between forewing and hindwing of a hovering dragonfly. *Acta Mechanica Sinica* **30**(6)**,** 787-799. https://doi.org/10.1007/s10409-014-0118-6.

**Hunt JCR, Wray AA and Moin P** (1988) Eddies, streams, and convergence zones in turbulent flows. In: *Proceedings of the 1988 summer program.* Stanford, CA.

**King JT, Kumar R and Green MA** (2018) Experimental observations of the three-dimensional wake structures and dynamics generated by a rigid, bioinspired pitching panel. *Physical Review Fluids* **3**(3)**,** 034701.

**Ko H, Girma A, Zhang Y, Pan Y, Lauder GV and Nagpal R** (2025) Beyond planar: fish schools adopt ladder formations in 3D. *Scientific Reports***,** in press.

**Kruyt JW, van Heijst GF, Altshuler DL and Lentink D** (2015) Power reduction and the radial limit of stall delay in revolving wings of different aspect ratio. *Journal of The Royal Society Interface* **12**(105)**,** 20150051. https://doi.org/10.1098/rsif.2015.0051.

**Kurt M, Eslam Panah A and Moored KW** (2020) Flow interactions between low aspect ratio hydrofoils in in-line and staggered arrangements. *Biomimetics* **5**(2)**,** 13.

**Kurt M and Moored K** (2018) Unsteady performance of finite-span pitching propulsors in side-by-side arrangements. In.

**Lehmann F-O** (2009) Wing–wake interaction reduces power consumption in insect tandem wings. *Experiments in fluids* **46,** 765-775.

**Li X, Gu J, Su Z and Yao Z** (2021) Hydrodynamic analysis of fish schools arranged in the vertical plane. *Physics of Fluids* **33**(12).

**Liu X, Hefler C, Fu J, Shyy W and Qiu H** (2021) Implications of wing pitching and wing shape on the aerodynamics of a dragonfly. *Journal of Fluids and Structures* **101,** 103208. https://doi.org/10.1016/j.jfluidstructs.2020.103208.

**Menzer A, Pan Y, Lauder GV and Dong H** (2025) Fish schools in a vertical diamond formation: Effect of vertical spacing on hydrodynamic interactions. *Physical Review Fluids* **10**(4)**,** 043104.

**Mittal R, Dong H, Bozkurttas M, Najjar F, Vargas A and Von Loebbecke A** (2008) A versatile sharp interface immersed boundary method for incompressible flows with complex boundaries. *J Comput Phys* **227**(10)**,** 4825-4852.

**Nan Y, Peng B, Chen Y, Feng Z and McGlinchey D** (2018) Can Scalable Design of Wings for Flapping Wing Micro Air Vehicle Be Inspired by Natural Flyers? *International Journal of Aerospace Engineering* **2018,** e9538328. https://doi.org/10.1155/2018/9538328.

**Nauen JC and Lauder GV** (2002) Hydrodynamics of caudal fin locomotion by chub mackerel, Scomber japonicus (*Scombridae*). *The Journal of Experimental Biology* **205**(12)**,** 1709-1724.

**Norberg RÅ** (1975) Hovering Flight of the Dragonfly Aeschna Juncea L., Kinematics and Aerodynamics. In Wu TYT, Brokaw CJ and Brennen C (eds), *Swimming and Flying in Nature: Volume 2.* Boston, MA: Springer US, 763-781.

**Ormonde PC, Kurt M, Mivehchi A and Moored KW** (2024) Two-dimensionally stable self-organisation arises in simple schooling swimmers through hydrodynamic interactions. *Journal of Fluid Mechanics* **1000,** A90.
**Pan Y, Dong H and Zhang W** (2021) An improved level-set-based immersed boundary reconstruction method for computing bio-Inspired underwater propulsion. In: *ASME 2021 Fluids Engineering Division Summer Meeting.* Virtual, Online.
**Pan Y, Wang J and Dong H** (2019) Study on the passive pitching mechanism of different forms of flapping motion in turning flight. In: *AIAA Aviation 2019 Forum.*
**Pan Y, Zhang W, Kelly J and Dong H** (2024) Unraveling hydrodynamic interactions in fish schools: a three-dimensional computational study of in-line and side-by-side configurations. *Physics of Fluids* **36**(8).
**Sambilay V** (1990) Interrelationships between swimming speed, caudal fin aspect ratio and body length of fishes. In.
**Standen E and Lauder GV** (2007) Hydrodynamic function of dorsal and anal fins in brook trout (Salvelinus fontinalis). *Journal of Experimental Biology* **210**(2)**,** 325-339.
**Sun M and Lan SL** (2004) A computational study of the aerodynamic forces and power requirements of dragonfly (Aeschna juncea) hovering. *Journal of Experimental Biology* **207**(11)**,** 1887-1901. https://doi.org/10.1242/jeb.00969.
**Triantafyllou GS, Triantafyllou MS and Grosenbaugh MA** (1993) Optimal thrust development in oscillating foils with application to fish propulsion. *Journal of Fluids and Structures* **7**(2)**,** 205-224.
**Usherwood JR and Ellington CP** (2002) The aerodynamics of revolving wings II. Propeller force coefficients from mayfly to quail. *Journal of Experimental Biology* **205**(11)**,** 1565-1576. https://doi.org/10.1242/jeb.205.11.1565.
**Van Buren T, Floryan D, Brunner D, Senturk U and Smits AJ** (2017) Impact of trailing edge shape on the wake and propulsive performance of pitching panels. *Physical Review Fluids* **2**(1)**,** 014702.
**Wainwright PC, Bellwood DR and Westneat MW** (2002) Ecomorphology of Locomotion in Labrid Fishes. *Environmental Biology of Fishes* **65**(1)**,** 47-62. https://doi.org/10.1023/A:1019671131001.
**Walker JA and Westneat MW** (2002) Performance limits of labriform propulsion and correlates with fin shape and motion. *Journal of Experimental Biology* **205**(2)**,** 177-187. https://doi.org/10.1242/jeb.205.2.177.
**Wang JK and Sun M** (2005) A computational study of the aerodynamics and forewing-hindwing interaction of a model dragonfly in forward flight. *The Journal of Experimental Biology* **208**(Pt 19)**,** 3785-3804. https://doi.org/10.1242/jeb.01852.
**Wang ZJ and Russell D** (2007) Effect of forewing and hindwing interactions on aerodynamic forces and power in hovering dragonfly flight. *Phys Rev Lett* **99**(14)**,** 148101. https://doi.org/10.1103/PhysRevLett.99.148101.
**Webb PW** (1994) The biology of fish swimming. *Mechanics and physiology of animal swimming* **4562**.
**Weihs D** (1989) Design Features and Mechanics of Axial Locomotion in Fish1. *American Zoologist* **29**(1)**,** 151-160. https://doi.org/10.1093/icb/29.1.151.
**Wu TY** (2011) Fish Swimming and Bird/Insect Flight. *Annual Review of Fluid Mechanics* **43**(1)**,** 25-58. https://doi.org/10.1146/annurev-fluid-122109-160648.
**Xie C and Huang W-X** (2015) Vortex interactions between forewing and hindwing of dragonfly in hovering flight. *Theoretical and Applied Mechanics Letters* **284**. https://doi.org/10.1016/j.taml.2015.01.007.
**Yeh PD and Alexeev A** (2016) Effect of aspect ratio in free-swimming plunging flexible plates. *Computers & Fluids* **124,** 220-225.
29

**Zhang W, Pan Y, Wang J, Di Santo V, Lauder GV and Dong H** (2023) An efficient tree-topological local mesh refinement on Cartesian grids for multiple moving objects in incompressible flow. *J Comput Phys*, 111983.

**Zhu Y and Breuer K** (2023) Flow-induced oscillations of pitching swept wings: Stability boundary, vortex dynamics and force partitioning. *Journal of Fluid Mechanics* **977,** A1.